\documentclass[runningheads]{llncs}

\usepackage{amssymb}
\usepackage{graphicx}
\usepackage{amsmath}
\usepackage{pgfplots}
\usepgfplotslibrary{groupplots}
\usepackage{array,multirow}

\newcommand{\STAB}[1]{\begin{tabular}{@{}c@{}}#1\end{tabular}}

\usepackage{url}
\usepackage{xcolor}
\usepackage{hyperref}
\makeatletter
\def\UrlAlphabet{%
      \do\a\do\b\do\c\do\d\do\e\do\f\do\g\do\h\do\i\do\j%
      \do\k\do\l\do\m\do\n\do\o\do\p\do\q\do\r\do\s\do\t%
      \do\u\do\v\do\w\do\x\do\y\do\z\do\A\do\B\do\C\do\D%
      \do\E\do\F\do\G\do\H\do\I\do\J\do\K\do\L\do\M\do\N%
      \do\O\do\P\do\Q\do\R\do\S\do\T\do\U\do\V\do\W\do\X%
      \do\Y\do\Z}
\def\UrlDigits{\do\1\do\2\do\3\do\4\do\5\do\6\do\7\do\8\do\9\do\0}
\g@addto@macro{\UrlBreaks}{\UrlOrds}
\g@addto@macro{\UrlBreaks}{\UrlAlphabet}
\g@addto@macro{\UrlBreaks}{\UrlDigits}
\makeatother

\begin{document}

\mainmatter  % start of an individual contribution

% first the title is needed
\title{Robust Retinal Vessel Segmentation from a Data Augmentation Perspective}

% a short form should be given in case it is too long for the running head
\titlerunning{Robust Retinal Vessel Segmentation from a Data Augmentation Perspective}

% the name(s) of the author(s) follow(s) next
%
% NB: Chinese authors should write their first names(s) in front of
% their surnames. This ensures that the names appear correctly in
% the running heads and the author index.
%
\author{Xu Sun\thanks{These authors contributed equally to this work.} \and
Huihui Fang\footnotemark[1]\and
Yehui Yang \and
Dongwei Zhu \and
Lei Wang  \and
Junwei Liu  \and
Yanwu Xu\thanks{Corresponding author.}}

\authorrunning{Sun et al.}
% First names are abbreviated in the running head.
% If there are more than two authors, 'et al.' is used.

\institute{Intelligent Healthcare Unit, Baidu Inc., Beijing, China\\\email{ywxu@ieee.org}}

\toctitle{Lecture Notes in Computer Science}
\tocauthor{Authors' Instructions}
\maketitle

\begin{abstract}
Retinal vessel segmentation is a fundamental step in screening, diagnosis, and treatment of various cardiovascular and ophthalmic
diseases. Robustness is one of the most critical requirements for practical utilization, since the test images may be captured using different
fundus cameras, or be affected by various pathological changes. We investigate this problem from a data augmentation perspective, with the
merits of no additional training data or inference time. In this paper,
we propose two new data augmentation modules, namely, channel-wise
random Gamma correction and channel-wise random vessel augmentation. Given a training color fundus image, the former applies random
gamma correction on each color channel of the entire image, while the
latter intentionally enhances or decreases only the fine-grained blood
vessel regions using morphological transformations. With the additional
training samples generated by applying these two modules sequentially,
a model could learn more invariant and discriminating features against
both global and local disturbances. Experimental results on both realworld and synthetic datasets demonstrate that our method can improve
the performance and robustness of a classic convolutional neural network
architecture. The source code is available at \url{https://github.com/PaddlePaddle/Research/tree/master/CV/robust_vessel_segmentation}.
\end{abstract}

\section{Introduction}

Retinal vessel segmentation plays a crucial role in computer-aided screening, diagnosis, and treatment of various cardiovascular and ophthalmic diseases such as stroke, diabetics, hypertension and retinopathy of prematurity~\cite{kanski2011clinical}. A substantial amount of work has been reported in the last two decades for automated detecting blood vessels in retinal fundus images. These algorithms can mainly be categorized into two groups: the unsupervised and supervised methods. The unsupervised methods rely on strong but intuitive priors of the blood vessel appearance~\cite{bankhead2012fast,fan2018hierarchical}, while the supervised methods utilize labelled datasets based on given features~\cite{fraz2012ensemble,soares2006retinal}. Among these algorithms,  supervised segmentation of blood vessels based on deep learning has reached new performance levels~\cite{araujo2019deep,fu2016deepvessel,zhang2019attention}. 

Despite architectural advances based on deep learning have led to enormous progress at segmenting vessels in curated datasets, their ability to generalize to new situations is rarely studied. In contrast, the generalization ability, which refers to robustness, is an important factor for algorithms performance. To improve robustness, there exist two issues which need special attention. First, in the real world context of retinal fundus image analysis, the input images may come from different kinds of digital fundus camera systems. Since the tonal quality of a fundus image is affected by the characteristics of these systems~\cite{tyler2009characteristics}, models fitting well to datasets collected from a specific class of fundus camera might fail to generalize to those captured from other types of machines. Second, for retinal vessel segmentation models to be adopted in practice, they also need to be robust on pathological changes, especially on those not included during the training stage. 

These issues can be alleviated by different strategies. Image pre-processing techniques like contrast limited adaptive histogram equalization try to shrink the difference among samples by redistributing their pixel values, However, they only lead to limited improvement yet require additional inference time. Domain adaptation, on the other hand, learns to adapt models between domains. But more data from the target domain are needed to retrain the models when encountering a new circumstance. In contrast, data augmentation methods, which includes input transforms that the model should be invariant against, show great merits of without requiring any extra training data or inference time. Motivated by that, in this paper we investigate the robust retinal vessel segmentation problem from a data augmentation perspective.

Our method consists of two novel data augmentation modules, {\it i.e.}, channel-wise random gamma correction and channel-wise random vessel augmentation, for training robust retinal vessel segmentation models. The former aims at varying the tonal quality of the whole image, while the latter only focuses on augmenting the visual appearance of retinal vessels. By doing so, the models are able to learn more representative features regardless of both global and local variations. The experimental results on three real world datasets suggest that the proposed method significantly increases the robustness on samples that are captured by different camera systems and/or affected by diverse pathological changes. Furthermore, we also conduct a thorough set of synthetic datasets to demonstrate that our augmentation scheme achieves reduced sensitivity to the variations of image brightness, contrast and saturation.

%To illustrate the effectiveness of our methods, we conduct a thorough set of ablation study experiments. The results suggest that our methods lead to reduced sensitivity to the variations of image brightness, contrast and saturation. Moreover, by applying models trained on the DRIVE training set to various databases, we find that the proposed methods significantly increase the robustness on samples that are corrupted by severe color distortion, captured by different camera systems and/or affected by diverse pathological changes.

%\begin{figure}[!t]
%	\centering
%	\includegraphics[width=0.95\linewidth]{framework.pdf}
%	\caption{Architecture of the proposed method for OD and OC segmentation/localization, where purple and magenta regions denote OD and OC respectively.}
%	\label{fig:framework}
%\end{figure}

\begin{figure}[t!]
\centering
\includegraphics[width=0.85\textwidth]{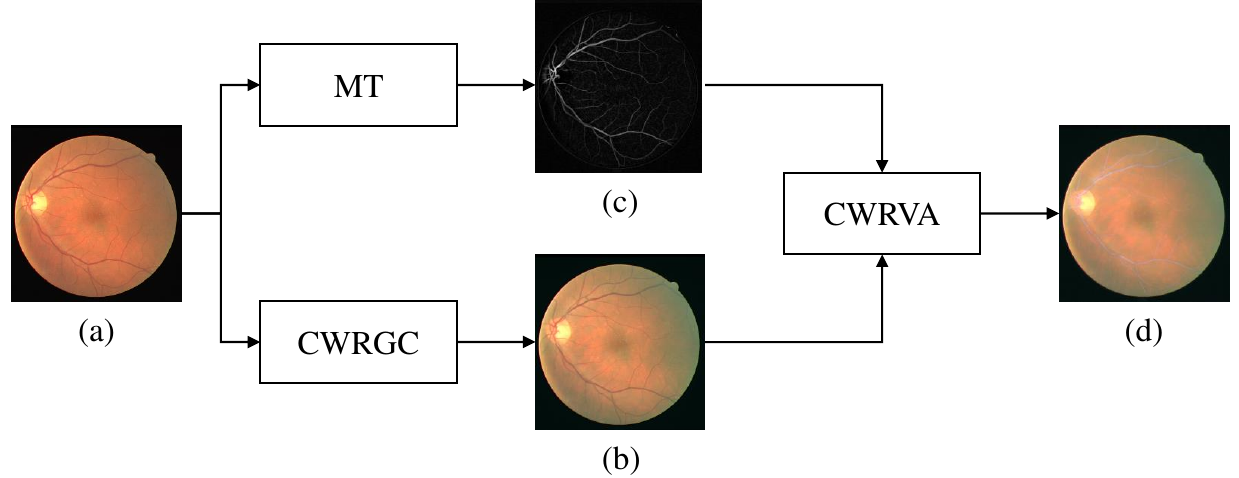}\\
%\begin{tabular}{ccc}
%  \includegraphics[height=40mm]{21_training-mini.pdf} &
%  \includegraphics[height=40mm]{21_training_rgc_2-mini.pdf} &
%  \includegraphics[height=40mm]{21_training_va-mini.pdf}\\
%  (a) & (b) & (c)\\
%\end{tabular}
\caption{Illustration of the proposed data augmentation scheme. (a) Original image from the DRIVE~\cite{staal2004ridge} training set. (b) Sample image augmented via channel-wise random gamma correction (CWRGC). (c) A rough vessel map generated by morphological transformation (MT). (d) Sample image augmented via channel-wise random vessel augmentation (CWRVA).}
\end{figure}

\section{Methodology}

In this section, we present a novel scheme to improve the robustness of retinal vessel segmentation, which comprises two data augmentation modules that increase the global and local invariance, respectively. 
Fig. 1 illustrates the process of virtual sample generation through the proposed data augmentation method.

%In this section, we present a novel scheme to enhance the robustness of retinal vessel segmentation models by including input transformations that the models should be invariant against. Fig. 1 illustrates how virtual examples are created through the proposed data augmentation method.

\subsection{Channel-wise Random Gamma Correction (CWRGC)}

Gamma correction is a nonlinear operation used to encode and decode luminance or tristimulus values, and has been widely used as a image preprocessing step in automated vessel segmentation systems. Unlike current approaches which employ gamma correction in the HSV (Hue, Saturation, value) color space~\cite{liskowski2016segmenting,zhou2017color}, we suggest to apply it directly in the RGB (Red, Green, Blue) color space. And, different from the preprocessing method to make the spatial distribution of training test samples more consistent, we use the data augmentation method to increase the diversity of sample distribution, so that the model learning can overcome the interference of task-independent features and learn more effective features. In particular, a simple yet effective data augmentation technique, termed channel-wise random gamma correction, is developed. This method is formulated as
\begin{equation}
\label{eq:cwrgc}
\widehat V_i = V_i^{\gamma_i}
\end{equation}
where $\widehat V_i$ and $V_i^{\gamma_i}$ represent the intensity of the image before and after transformation, respectively. $\gamma_i > 0$ is the correction value, and subscript $i \in \{\text{R, G, B}\}$ denotes the corresponding red, green, or blue channel. By varying $\gamma_i$ randomly, virtual examples covering a wide range of tonal quality can be created at the training stage. Fig. 1 (b) shows one of the generated images.

\subsection{Channel-wise Random Vessel Augmentation (CWRVA)}

Different from the first method that transforms the whole fundus images, our second method only focuses on blood vessel regions. This can be achieved by taking advantages of existing unsupervised methods, as they are able to provide the rough vessel maps without requiring annotations. To be specific, morphological transformation~\cite{leandro2001blood} is used here due to its simplicity of implementation and effectiveness in practice. When the structuring element used in the morphological opening is orthogonal to the vessel direction and longer than the vessel width, it will eradicate a vessel or part of it. Based on this observation, morphological transformation is defined as follows
\begin{equation}
\label{eq:tophat}
I_{\text{th}}^{\theta} = I - (I \circ S_e^{\theta}),
\end{equation}
\begin{equation}
\label{eq:tophatsum}
I_{S_{\text{th}}} = \sum_{\theta \in A}I_{\text{th}}^{\theta}
\end{equation}
where $I_{\text{th}}^{\theta}$ is the top-hat transformed image, $I$ is the image to be processed, $\circ$ is the opening operation, $S_e$ ia the structuring element, and $\theta \in A$ is the angular rotation equally distributed in $[0, \pi)$ .

Given the top-hat transformed image, the blood vessel attention map for each color channel of the fundus image can then be obtained by
\begin{equation}
\label{eq:amap}
M_i = \mathbb{N}(I_{S_{\text{th}}}) \cdot \lambda_i
\end{equation}
where $\mathbb{N}(\bf x)$ is a normalization function which scales and shifts the input array $\bf x$ so that the minimum and maximum value of $\bf x$ are 0 and 1, respectively, and $\lambda_i \in [0, 1]$ is a random decay coefficient with $i \in \{\text{R, G, B}\}$.

The proposed channel-wise random vessel augmentation is formulated as
\begin{equation}
\label{eq:cwrva}
\widetilde V_i = V_i \cdot (1 - M_i) + M_i \cdot 255
\end{equation}
where $\widetilde V_i$ and $V_i$ denote the intensity values of the image before and after vessel augmentation, respectively. Virtual images with various visual effect can be generated through changing $\lambda_i$ in equation (\ref{eq:amap}). A typical example is shown in Fig. 1 (c).

\section{Experiments}

To evaluate the effectiveness of our method, a thorough set of ablation study experiments are conducted. The first experiment is performed on three real world datasets to show how our method impacts robustness on testing images collected by a different fundus camera and/or affected by different pathological changes. Furthermore, we also utilize synthetic datasets to investigate the sensitivity of different models to the variations of image brightness, contrast and saturation.

\subsection{Experiments Setup}

We adopt the U-Net architecture~\cite{ronneberger2015u} in our experiments due to its popularity in medical image analysis community and formation of the basis for most of the recent architectural advances at segmenting retinal vessels~\cite{gu2019net,wang2019dual}. In particular, we replace its feature encoder module with the pretrained ResNet-50, remaining the first five feature extraction blocks without the global averaging pooling layer and the fully connected layers.

We employ random horizontal flip and random vertical flip with a probability of 50\% as the basic data augmentation strategy (BS). In addition, two commonly used randomized data augmentation methods in literature are also implemented for comparison:
\begin{itemize}
  \item RGN: Disturb the intensity of the red, green and blue channels by adding Gaussian noise with mean of 0 and standard deviation of 20.
  \item SVGC: Gamma correction of Saturation and Value (of the HSV color space) by raising pixels to a power in [0.25, 4].
\end{itemize}
In our experiments, $\gamma_i$ in equation (\ref{eq:cwrgc}) for the channel-wise random gamma correction (CWRGC) is randomly selected from $[0.33, 3]$, and $\lambda_i$ in equation (\ref{eq:amap}) for the channel-wise random vessel augmentation (CWRVA) is randomly picked in $[0, 1]$.
All the models are trained on the DRIVE~\cite{staal2004ridge} training set using a publicly available library\footnote{https://github.com/PaddlePaddle/PaddleSeg}. We use the ``step-scaling'' method provided in the library to resize the input images to $640 \times 640$, setting the scaling factor range from 0.75 to 1.25 with a step of 0.25. We use adam as the optimizer. The learning rate is initially set to be 0.005 and then decays following the ``poly'' policy with a power of 0.9.  Instead of training all parameters from scratch, we fine-tune the network end-to-end from an ImageNet pre-trained model. We integrate both dice loss and binary cross entropy loss to train all models for 3000 epoches.

%A thorough set of ablation study experiments are conducted in this section. We employ random horizontal flip and random vertical flip with a probability of 50\% as the basic data augmentation strategy (BS) in our experiments.

Following previous work, the retinal vessel segmentation results are evaluated quantitatively by the area under the receiver operation characteristic curve (AUC), accuracy (ACC), specificity (SP), sensitivity (SE), and F1-score (F1). However, we mainly focus on AUC and F1 when comparing the performance of different methods as they are more reliable for evaluating binary classifiers (say, to classify if a pixel belongs to vessels or not)~\cite{japkowicz2011evaluating}. In particular, when we conclude that one method outperforms another, we mean that it achieves both the higher AUC and F1 if without stating which metrics are used.

\subsection{Generalization Across Different Datasets}\begin{table}[t!]
	\caption{Performance comparison on three Datasets.}
	\label{tab:qr}
	\begin{center}
		\begin{tabular}{|c|l||c|c|c|c|c|}
			\hline
			\;\;\;\;\;\; & Method & AUC & ACC & SP & SE & F1\\
			\hline
			\hline
            \multirow{6}{*}{\STAB{\rotatebox[origin=c]{90}{DRIVE}}}
			&BS                 & \;\;0.9755\;\; & \;\;0.9531\;\; & 0.9750 & \;\;0.8055\;\; & \;\;0.8126\;\;\\
			%\hline
            &BS + RGN           & 0.9769 & 0.9531 & 0.9746 & 0.8088 & 0.8131 \\
			%\hline
            &BS + SVGC           & 0.9772 & 0.9540 & {\bf 0.9752} & 0.8108 & 0.8167 \\
            \cline{2-7}
			&BS + CWRGC         & 0.9777 & 0.9539 & \;\;0.9744\;\; & 0.8150 & 0.8174\\
			%\hline
			&BS + CWRVA         & 0.9783 & {\bf 0.9545} & 0.9741 & 0.8225 & 0.8205\\
			%\hline
			&BS + CWRGC + CWRVA\;\; & {\bf 0.9788} & {\bf 0.9545} & 0.9741 & {\bf 0.8227} & {\bf 0.8209}\\
			\hline
            \hline
            \multirow{6}{*}{\STAB{\rotatebox[origin=c]{90}{STARE}}}
			&BS                 & \;\;0.9287\;\; & \;\;0.9334\;\; & 0.9512 & \;\;0.7103\;\; & \;\;0.6699\;\;\\
			%\hline
            &BS + RGN           & 0.9147 & 0.9556 & 0.9812 & 0.6157 & 0.6273 \\
			%\hline
            &BS + SVGC           & 0.9288 & 0.9602 & {\bf 0.9837} & 0.6500 & 0.6769 \\
            \cline{2-7}
			&BS + CWRGC         & 0.9892 & 0.9676 & \;\;0.9738\;\; & 0.8902 & 0.8056\\
			%\hline
			&BS + CWRVA         & 0.9771 & 0.9665 & 0.9816 & 0.7711 & 0.7652\\
			%\hline
			&BS + CWRGC + CWRVA\;\; & {\bf 0.9893} & {\bf 0.9683} & 0.9745 & {\bf 0.8908} & {\bf 0.8082}\\
			\hline
            \hline
            \multirow{6}{*}{\STAB{\rotatebox[origin=c]{90}{CHASE-DB1}}}
			&BS                 & \;\;0.8794\;\; & \;\;0.9344\;\; & 0.9825 & \;\;0.2989\;\; & \;\;0.3768\;\;\\
			%\hline
            &BS + RGN           & 0.9165 & 0.9394 & 0.9715 & 0.5159 & 0.5279 \\
			%\hline
            &BS + SVGC           & 0.9258 & 0.9465 & {\bf 0.9893} & 0.3848 & 0.4841 \\
            \cline{2-7}
			&BS + CWRGC         & 0.9812 & {\bf 0.9623} & \;\;0.9702\;\; & 0.8565 & 0.7563\\
			%\hline
			&BS + CWRVA         & 0.9555 & 0.9522 & 0.9772 & 0.6230 & 0.6385\\
			%\hline
			&BS + CWRGC + CWRVA\;\; & {\bf 0.9838} & 0.9612 & 0.9673 & {\bf 0.8818} & {\bf 0.7565}\\
			\hline
		\end{tabular}
	\end{center}
\end{table}

\begin{figure}[t!]
\centering
\includegraphics[width=0.85\textwidth]{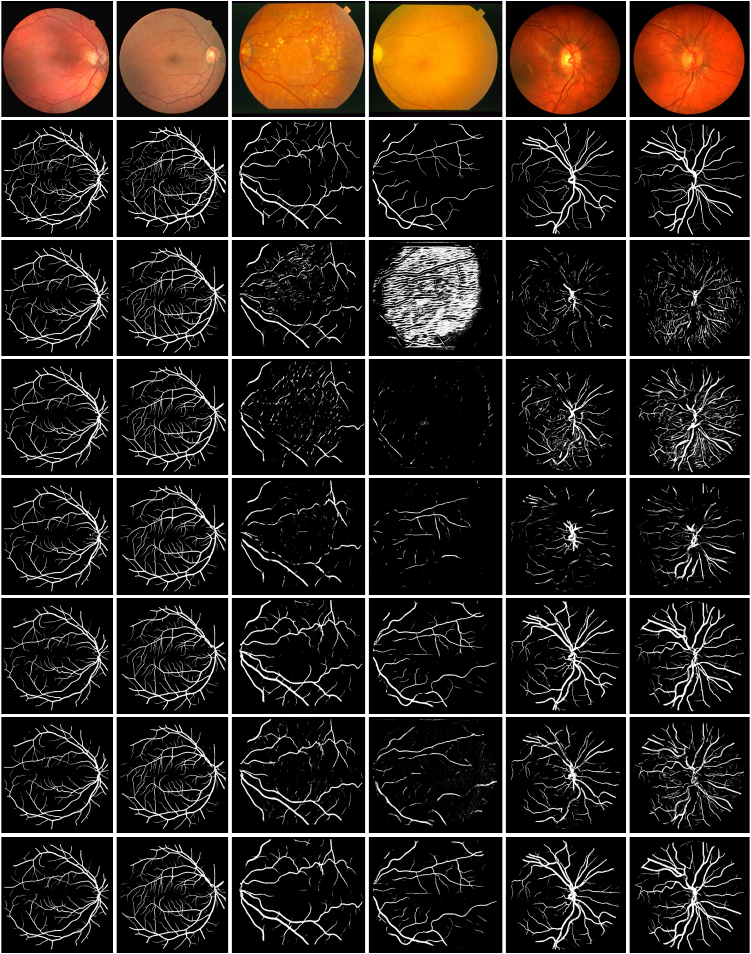}
%(a) first & (b) second \\[1pt]
% \includegraphics[width=65mm]{it} &   \includegraphics[width=65mm]{it} \\
%(c) third & (d) fourth \\[6pt]
%\multicolumn{2}{c}{\includegraphics[width=65mm]{it} }\\
%\multicolumn{2}{c}{(e) fifth}
\caption{Sample results on three different datasets. From up to bottom: input image, ground truth and predictive results of BS, BS+RGN, BS+SVGC, BS+CWRGC, BS+CWRVA, and BS+CWRGC+CWRVA. From left to right: sample images from DRIVE, DRIVE, STARE, STARE, CHASE-DB1, CHASE-DB1.}
\label{fig:sr}
\end{figure}

To validate how our augmentations impact robustness in a realistic setting, models trained on the DRIVE training set are applied to three datasets:
\begin{itemize}
  \item {\bf Testing set of DRIVE}~\cite{staal2004ridge}: the fundus images are captured from the same digital fundus camera system.
  %\item {\bf Testing set of DRIVE-GRAY}: the same images as the testing set of DRIVE except that all of them are converted to gray images.
  \item {\bf Full set of STARE}~\cite{hoover2000locating}: the provided images come by a different type of fundus camera, and contain more kinds of pathological changes
  \item {\bf Full set of CHASE-DB1}~\cite{owen2009measuring}: the images are captured by another type of fundus machine which has a smaller field of view.
\end{itemize}
From the evaluation results shown in Table \ref{tab:qr}, we can observe that: 1) BS+CWR-GC+CWRVA achieves the best results in all datasets; 2) BS+CWRGC and BS+CWRVA outperform BS, BS+RGN and BS+SVGC in all datasets. 3) BS performs worst on DRIVE, DRIVE-GRAY and CHASE-DB1, but outperforms BS+RGN in STARE; 4) the performance of BS+RGN and BS+SVGC degenerates significantly on STARE and CHASE-DB1; 5) although CWRVA works slightly better than CWRGC in the DRIVE testing set, such superiority fails to generalize to other datasets; 6) SVGC achieves the highest SP in all testing set, at the expense of getting a much lower SE when comparing to BS+CWRGC, BS+CWRVA, and BS+CWRGC+CWRVA; 7) CWRGC outperform SVGC in all datasets in terms of AUC and F1. This experiment shows that in RGB space is indeed significantly better than in HSV space due that directly applied to RGB space can be targeted to optimize the hue change problem. Fig. \ref{fig:sr} shows some visual examples of the segmentation results. The results on real world datasets suggest that the proposed method possess the significant generalization improvement to samples captured by different camera systems and/or affected by diverse pathological changes. This is due to the different camera systems resulting in the various image hues, while gamma correction is good at generating augmentation images with multiple different hues to improve the robustness of the model. In addition, pathological changes are mainly reflected in the local changes of fundus image texture and vascular region, and random vessel augmentation could increase the variation diversity of vascular region, thus improving the robustness of the model in the region with pathological changes.

\subsection{Robustness to Brightness, Contrast and Saturation}

\begin{figure}[t!]
%\pgfplotsset{compat=1.3}
\centering
\includegraphics[width=0.85\textwidth]{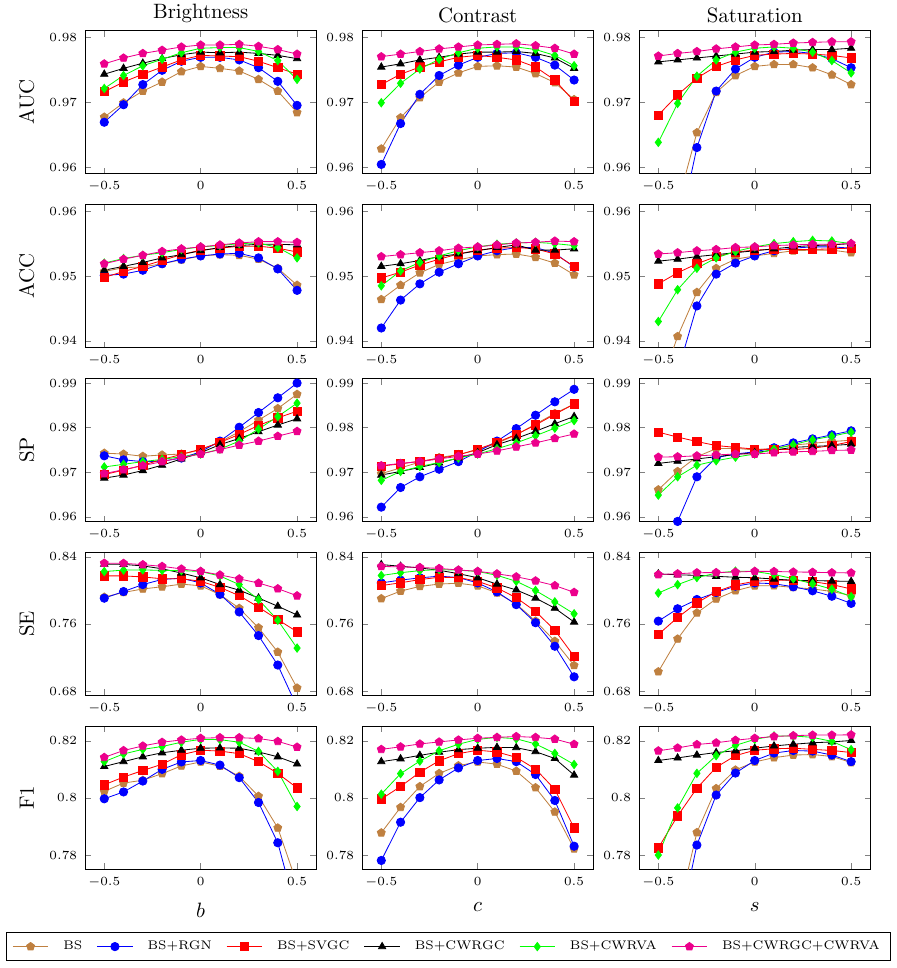}
\caption{Evaluation results to illustrate robustness of different methods to variations of brightness, contrast and saturation}
\label{fig:bcs}
\end{figure}

In order to investigate a model's robustness on the more complex situation, the virtual dataset are employed. Thus, three image processing functions are respectively introduced to adjust the brightness, contrast and saturation of a color fundus image. Let $V$ be the input image in RGB space, ${\mathbb G}(V)$ be the function to convert the input image from RGB space to gray space, and ${\mathbb M}(V)$ be the mean function, the brightness jitter, contrast jitter and saturation jitter can then be defined, respectively,  as
\begin{eqnarray}
  \label{eq:brightnessjitter}
  {\mathbb B}(V) &=& V \cdot (1 - b),\\
  \label{eq:brightnessjitter}
  {\mathbb C}(V) &=& V \cdot (1 - c) + {\mathbb M}({\mathbb G}(V)) \cdot c,\\
  \label{eq:brightnessjitter}
  {\mathbb S}(V) &=& V \cdot (1 - s) + {\mathbb G}(V) \cdot s
\end{eqnarray}
where $b \in [-1, 1]$ is brightness jitter ratio, $c \in [-1, 1]$ is the contrast jitter ratio, and $s \in [-1, 1]$ is saturation jitter ratio. The output values of these functions are all limited to $[0, 255]$.

By respectively varying $b$, $c$ and $s$ from -0.5 to 0.5 with a step of 0.1, we construct 30 more datasets with different degree of brightness, contrast and saturation based on the testing set of DRIVE. The evaluation results of different data augmentation strategies on these datasets are shown in Fig. \ref{fig:bcs}. It can be obviously seen that : 1) models trained with the proposed methods are less sensitive to the variations of image brightness, contrast and saturation than BS and BS+RGN in terms of all the five evaluation metrics; 2) BS+CWRGC+CWRVA method consistently achieves the best results for all sorts of settings. These results indicates the proposed algorithms lead to reduced sensitivity to these naturally occurring variations.

\section{Conclusion}
This paper investigates the practicability and robustness of retinal vessel segmentation from a data augmentation perspective, with the advantages of not requiring extra training data or inference time. Our method comprises two new data augmentation modules to increase the performance and robustness of models learned. The channel-wise random gamma correction module aims at covering a wide range of tonal quality of the global image, while the channel-wise random vessel augmentation module focuses on diversifying the local visual appearance of the retinal vessels only. The proposed methods achieve excellent results on both real-world and virtual datasets. Experimental results on various real-world public datasets show that the proposed method could consistently stabalize the segmentation performance on samples captured by different cameras or affected by various pathological changes. Moreover, by conducting synthetic databases, we also observe that the proposed method is less sensitive to the variations of image brightness, contrast, and saturation. In the future, we plan to explore more general techniques for robust automated image analyzing systems.

\bibliographystyle{splncs}
\bibliography{rvs}

\begin{thebibliography}{10}
\providecommand{\url}[1]{\texttt{#1}}
\providecommand{\urlprefix}{URL }

\bibitem{araujo2019deep}
Ara{\'u}jo, R.J., Cardoso, J.S., Oliveira, H.P.: A deep learning design for
  improving topology coherence in blood vessel segmentation. In: International
  Conference on Medical Image Computing and Computer-Assisted Intervention. pp.
  93--101. Springer (2019)

\bibitem{bankhead2012fast}
Bankhead, P., Scholfield, C.N., McGeown, J.G., Curtis, T.M.: Fast retinal
  vessel detection and measurement using wavelets and edge location refinement.
  PloS one  7(3) (2012)

\bibitem{fan2018hierarchical}
Fan, Z., Lu, J., Wei, C., Huang, H., Cai, X., Chen, X.: A hierarchical image
  matting model for blood vessel segmentation in fundus images. IEEE
  Transactions on Image Processing  28(5),  2367--2377 (2018)

\bibitem{fraz2012ensemble}
Fraz, M.M., Remagnino, P., Hoppe, A., Uyyanonvara, B., Rudnicka, A.R., Owen,
  C.G., Barman, S.A.: An ensemble classification-based approach applied to
  retinal blood vessel segmentation. IEEE Transactions on Biomedical
  Engineering  59(9),  2538--2548 (2012)

\bibitem{fu2016deepvessel}
Fu, H., Xu, Y., Lin, S., Wong, D.W.K., Liu, J.: Deepvessel: Retinal vessel
  segmentation via deep learning and conditional random field. In:
  International conference on medical image computing and computer-assisted
  intervention. pp. 132--139. Springer (2016)

\bibitem{gu2019net}
Gu, Z., Cheng, J., Fu, H., Zhou, K., Hao, H., Zhao, Y., Zhang, T., Gao, S.,
  Liu, J.: Ce-net: context encoder network for 2d medical image segmentation.
  IEEE transactions on medical imaging  38(10),  2281--2292 (2019)

\bibitem{hoover2000locating}
Hoover, A., Kouznetsova, V., Goldbaum, M.: Locating blood vessels in retinal
  images by piecewise threshold probing of a matched filter response. IEEE
  Transactions on Medical imaging  19(3),  203--210 (2000)

\bibitem{japkowicz2011evaluating}
Japkowicz, N., Shah, M.: Evaluating learning algorithms: a classification
  perspective. Cambridge University Press (2011)

\bibitem{kanski2011clinical}
Kanski, J.J., Bowling, B.: Clinical ophthalmology: a systematic approach.
  Elsevier Health Sciences (2011)

\bibitem{leandro2001blood}
Leandro, J.J., Cesar, J., Jelinek, H.F.: Blood vessels segmentation in retina:
  Preliminary assessment of the mathematical morphology and of the wavelet
  transform techniques. In: Proceedings XIV Brazilian Symposium on Computer
  Graphics and Image Processing. pp. 84--90. IEEE (2001)

\bibitem{liskowski2016segmenting}
Liskowski, P., Krawiec, K.: Segmenting retinal blood vessels with deep neural
  networks. IEEE transactions on medical imaging  35(11),  2369--2380 (2016)

\bibitem{owen2009measuring}
Owen, C.G., Rudnicka, A.R., Mullen, R., Barman, S.A., Monekosso, D., Whincup,
  P.H., Ng, J., Paterson, C.: Measuring retinal vessel tortuosity in
  10-year-old children: validation of the computer-assisted image analysis of
  the retina (caiar) program. Investigative ophthalmology \& visual science
  50(5),  2004--2010 (2009)

\bibitem{ronneberger2015u}
Ronneberger, O., Fischer, P., Brox, T.: U-net: Convolutional networks for
  biomedical image segmentation. In: International Conference on Medical image
  computing and computer-assisted intervention. pp. 234--241. Springer (2015)

\bibitem{soares2006retinal}
Soares, J.V., Leandro, J.J., Cesar, R.M., Jelinek, H.F., Cree, M.J.: Retinal
  vessel segmentation using the 2-d gabor wavelet and supervised
  classification. IEEE Transactions on medical Imaging  25(9),  1214--1222
  (2006)

\bibitem{staal2004ridge}
Staal, J., Abr{\`a}moff, M.D., Niemeijer, M., Viergever, M.A., Van~Ginneken,
  B.: Ridge-based vessel segmentation in color images of the retina. IEEE
  transactions on medical imaging  23(4),  501--509 (2004)

\bibitem{tyler2009characteristics}
Tyler, M.E., Hubbard, L., Boydston, K., Pugliese, A.: Characteristics of
  digital fundus camera systems affecting tonal resolution in color retinal
  images. The Journal of Ophthalmic Photography  31(1),  1--9 (2009)

\bibitem{wang2019dual}
Wang, B., Qiu, S., He, H.: Dual encoding u-net for retinal vessel segmentation.
  In: International Conference on Medical Image Computing and Computer-Assisted
  Intervention. pp. 84--92. Springer (2019)

\bibitem{zhang2019attention}
Zhang, S., Fu, H., Yan, Y., Zhang, Y., Wu, Q., Yang, M., Tan, M., Xu, Y.:
  Attention guided network for retinal image segmentation. In: International
  Conference on Medical Image Computing and Computer-Assisted Intervention. pp.
  797--805. Springer (2019)

\bibitem{zhou2017color}
Zhou, M., Jin, K., Wang, S., Ye, J., Qian, D.: Color retinal image enhancement
  based on luminosity and contrast adjustment. IEEE Transactions on Biomedical
  Engineering  65(3),  521--527 (2017)

\end{thebibliography}

%\begin{thebibliography}{4}
%
%\bibitem{jour} Smith, T.F., Waterman, M.S.: Identification of Common Molecular
%Subsequences. J. Mol. Biol. 147, 195--197 (1981)
%
%\bibitem{lncschap} May, P., Ehrlich, H.C., Steinke, T.: ZIB Structure Prediction Pipeline:
%Composing a Complex Biological Workflow through Web Services. In: Nagel,
%W.E., Walter, W.V., Lehner, W. (eds.) Euro-Par 2006. LNCS, vol. 4128,
%pp. 1148--1158. Springer, Heidelberg (2006)
%
%\bibitem{book} Foster, I., Kesselman, C.: The Grid: Blueprint for a New Computing
%Infrastructure. Morgan Kaufmann, San Francisco (1999)
%
%\bibitem{proceeding1} Czajkowski, K., Fitzgerald, S., Foster, I., Kesselman, C.: Grid
%Information Services for Distributed Resource Sharing. In: 10th IEEE
%International Symposium on High Performance Distributed Computing, pp.
%181--184. IEEE Press, New York (2001)
%
%\bibitem{proceeding2} Foster, I., Kesselman, C., Nick, J., Tuecke, S.: The Physiology of the
%Grid: an Open Grid Services Architecture for Distributed Systems
%Integration. Technical report, Global Grid Forum (2002)
%
%\bibitem{url} National Center for Biotechnology Information, \url{http://www.ncbi.nlm.nih.gov}
%
%\end{thebibliography}

%\section*{Appendix: Springer-Author Discount}
%
%LNCS authors are entitled to a 33.3\% discount off all Springer
%publications. Before placing an order, the author should send an email,
%giving full details of his or her Springer publication,
%to \url{orders-HD-individuals@springer.com} to obtain a so-called token. This token is a
%number, which must be entered when placing an order via the Internet, in
%order to obtain the discount.

\end{document}